# Exploring the Impact of Ions on Oxygen K-Edge X-ray Absorption Spectroscopy in NaCl Solution using the GW-Bethe-Salpeter-Equation Approach

Fujie Tang[*], Kefeng Shi, Xifan Wu.

Department of Physics, Temple University, Philadelphia, PA 19122, USA

*Corresponding author: fujie.tang@temple.edu

**Abstract**

X-ray absorption spectroscopy (XAS) is a powerful experimental tool to probe the local structure in materials with the core hole excitations. Here, the oxygen K-edge XAS spectra of the NaCl solution and pure water are computed by using a recently developed GW-BSE approach, based on configurations modeled by path-integral molecular dynamics with the deep-learning technique. The neural network is trained on *ab initio* data obtained with SCAN density functional theory. The observed changes in the XAS features of the NaCl solution, compared to those of pure water, are in good agreement between experimental and theoretical results. We provided detailed explanations for these spectral changes that occur when NaCl is solvated in pure water. Specifically, the presence of solvating ion pairs leads to localization of electron-hole excitons. Our theoretical XAS results support the theory that the effects of the solvating ions on the H-bond network are mainly confined within the first hydration shell of ions, however beyond the shell the arrangement of water molecules remains to be comparable to that observed in pure water.

## I. Introduction

The interaction between hydrated NaCl ions and the hydrogen-bond (H-bond) network of water molecules in the salt solution is of fundamental importance for a broad range field such as biochemistry[1,2], geologic processes[3–5], and atmospheric chemistry[6]. For example, the Cl$^-$ anion belongs to the Hofmeister series of ions[7–9], known for their important roles on the modulation of the H-bonding network among water molecules within protein aqueous solutions, which referred to as the ability to either salt out or salt in proteins. Moreover, the Na$^+$ ion channel is a crucial cation-selective channel in the cell membrane, which allows for the passage of ions from one side of the membrane to the other[1,2]. All these various functionalities occur in the aqueous environments. Therefore, revealing the molecular structure of solvated ions is the central topic in understanding these processes of interest[10–15]. It is generally thought that the presence of ions can modify the H-bond network of water molecules, which are being intensively studied for decades by various experimental methods such as X-ray and neutron scattering[10,16–19] as well as X-ray absorption (XAS) spectroscopies[20–22]. These scattering experimental techniques provide structural information about the hydration ions, such as the Cl-O and Na-O radial distribution functions (RDFs), $g_{Cl-O}(r)$ and $g_{Na-O}(r)$. Based on these scattering and diffraction experiments, the referred structural information obtained at varying concentrations displayed a remarkable similarity to the behavior observed in pure water subjected to incremental pressure[23–25]. This seems to indicate that the H-bond network of water molecules in NaCl solution is distorted in a similar manner as it is in the pure water under high pressure[23–25]. In accordance with the aforementioned hypothesis, the presence of ions in salt solutions not only distorts the short-range ordering of the H-bond network but also alters the long-range ordering, deviating from the H-bond network observed in pure water[23–25]. The interpretation of these scattering and diffraction experiments challenges the idea that NaCl ions have a small effect among the Hofmeister series[26,27].

Complementary to the thermally averaged structural information obtained from the scattering experiments, the oxygen K-edge XAS measurement carries out an instantaneous local fingerprint of water structure, because the time scale of the electron-hole excitation process is much shorter than that of the molecular relaxation in the XAS process[28]. Since the different types of excitations associated with features at pre-edge, main-edge, and post-edge of the XAS spectra give the short-range, intermediate-range, and long-range structural information[29,30], respectively, XAS can probe the delicate change of the molecular structures of H-bond network induced by the additional ions pair in the NaCl solution[31–33]. The experimental XAS of salt solutions indeed shows specific enhancements in the pre-edge, main-edge peaks, and slight decrease in the post-edge peak compared with the XAS of pure water[31–34]. Therefore, the above controversy could be reconcilable through the XAS measurement and its interpretation. However, such a comprehensive elucidation regarding the interaction between the ions and their surrounding H-bond network as well as its connections to the spectral feature remain unclear[31–33].

The unambiguous assignments of the spectral feature and the underlying H-bond structures in the NaCl solution can be achieved by using theoretical XAS calculation, which requires the accurate modeling for both molecular structures and electron-hole interaction. By extracting the representative snapshots from the equilibrated molecular dynamics simulation trajectories, one can calculate the XAS spectra with the knowledge of electronic structure of the excited holes and electrons. For many years, the accurate modeling of water and salt solution is a challenging task using *ab initio* molecular dynamics (AIMD) simulation, as it has been validated that the nuclear quantum effects (NQEs) accessed by using path-integral molecular dynamics (PIMD) simulation, the van der Waals interaction, and the exact exchange are the key factors[35–44]. To include such fine effects, one needs to adopt a non-local exchange-correlation functional which is in the high rung of the metaphorical Jacob's ladder[45]. Recently, by using SCAN functional and including the NQEs in the simulation, we have reached a great accuracy in molecular structures of the water and NaCl solution compared to the experiment[27,43,46,47]. Moreover, by using the state-of-the-art molecular analysis with the SCAN functional, we have successfully reproduced the structure factors $S_{OO}(Q)$ of salt water, and discovered that the influence of solvated ions is mostly confined to the first hydration shells, beyond which the oxygen radial-distribution function remains relatively unchanged compared to pure water[27].

On the other hand, the first-principles calculation for XAS spectra is computational formidable for decades as the difficulties for a large supercell in solving the Bethe-Salpeter equation (BSE), where the many-body effects of the electron-hole interaction are considered[22,48]. As such, many calculations[31–33,49–58], to date, employ the frozen core-hole approximations, where the dynamic effects of core-hole are ignored, and the two-body correlated electron–hole excitation is simplified as an equivalent one-particle excitation process. By employing the density functional theory (DFT), Saykally and coworkers[31,33,34] demonstrated that spectral change induced by the ions indicated the distortion of the H-bond network is confined within the first solvation shell of the ions. Galli and coworkers[32] calculated the XAS spectra of salt water with the frozen core-hole methods to investigate the effects that $Mg^{2+}$, $Ca^{2+}$, and $Na^+$ have on the XAS spectrum of the respective solutions and establish the correlation between molecular structures and spectral feature. Recently, we found that to accurately reproduce the XAS spectra of liquid water and its aqueous solution, the missing many-body effects[29], including: 1) the quasiparticle wavefunctions treatment which beyond density functional theory approximations, accounts for the dynamics of quasiparticles, and 2) the dynamic screening as well as renormalization effects due to the continuum of valence-level excitations are of notable importance. Therefore, despite the continuing theoretical efforts in modeling XAS spectra of salt solution[31–33], the investigation from an accurate spectroscopic view about the effects of the solvated ions on the H-bond network of water remains to be done.

To address the above issues, we will apply a newly developed GW-BSE approach to simulate the XAS spectra of salt solution. This efficient GW-BSE approach[29] has been proven to allow us to calculate the XAS of liquid water from *ab initio* with unprecedented accuracy for both the relative energies of the pre-edge, main-edge, and post-edge features and their spectral line shape. In this approach[29], sophisticated techniques are developed to address molecular structures by employing path-integral deep potential molecular dynamics (PI-DPMD) at the higher rung of the Jacob's ladder. Additionally, electronic structures are tackled by employing quasi-particle wavefunctions computed within the GW approximation, while also accounting for renormalization effects attributed to the continuum of valence-level excitations. By employing this approach, we have successfully replicated the subtle spectral variances observed in the pre-edge, main-edge, and post-edge peaks between the spectra obtained from the NaCl solution and pure water. These effects are typically ascribed to the presence of substituted ion pairs within the tetrahedral H-bond network of water molecules, resulting in the localized charge distributions of excitons. We gave a set of detailed assignments and explanations for the three pre-edge, main-edge, post-edge peaks and their molecular origins in this work.

The rest of this paper is organized as follows. The computational details for MD simulation, the criteria for picking snapshots used for the XAS calculations, and the details of GW-BSE XAS calculation are given in Sec. II. The theoretical XAS spectra of NaCl solution, the impact of NQEs on XAS spectra, and the detailed analysis of the pre-edge, main-edge peaks are given in Sec. III. Finally, we conclude our work with a short summary in the Sec. IV.

## II. Methods

### A. Details of the Molecular Dynamics Simulation

The molecular structures for NaCl solution and pure water used in our XAS calculation were extracted from the trajectories of PI-DPMD and DPMD simulations, where the neural network potential was trained by using DFT data based on the strongly constrained and appropriately normed (SCAN) XC functional[27,59]. Our PI-DPMD and DPMD simulations were performed in the canonical ensemble (*NVT*) at $T = 300$ K with periodic boundary conditions. In our simulated XAS spectra of NaCl solution, we consider 2 M solution as an example, which is corresponding to one

NaCl ion pair and 29 water molecules in our chosen cell size. 32 water molecules are contained in our simulation cell for pure water. The cell size is adjusted to have the same density as the experimental value, with 9.529 Å × 9.529 Å × 9.529 Å for both pure water and NaCl solution. The choice of the current concentration in simulation gives a reasonable ratio between ions and water molecules in a limited-size simulation cell. The temperature was controlled by using an eight-bead ring polymer with a color noise path-integral generalized Langevin equation thermostat (i.e. PIGLET)[60] in PI-DPMD. All the PI-DPMD and DPMD simulations used well equilibrated ~500 ps long trajectories.

To compute the XAS spectra of NaCl salt solution and pure water, we need to average over snapshots of the molecular dynamics simulation trajectories. However, due to the heavy computational cost of the GW-BSE calculation, we could only use a limited number of snapshots to calculate the spectra. In order to sample the most representative snapshots, we used a similar approach reported in the previous paper[29], we assigned to each snapshot $i$ as a score function $f(i)$ that measures the deviation of the structure at snapshot $i$ relative to the average structure in a trajectory. The score function $f(i)$ defined as:

$$f(i) = \sum_{k=1}^{8} |X_k^i - \bar{X}_k|/|\bar{X}_k|.$$

Here $X_k^i (k = 1, \ldots, 8)$ are the descriptors of intra-molecular structures, structures of H-bond network, and thermodynamic properties at snapshot $i$. Specifically, $X_{k=1}^i$ is the average proton transfer distance $\delta = r(O \cdots H) - r(OH)$, $X_{k=2}^i$ is the average covalent bond length $r(OH)$, $X_{k=3}^i$ is the average number of H-bonds, $X_{k=4}^i$ is the average H-bond length, $X_{k=5}^i$ is the average O-O nearest neighbor distance, $X_{k=6}^i$ is the average O-Na nearest neighbor distance, $X_{k=7}^i$ is the average O-Cl nearest neighbor distance, $X_{k=8}^i$ is the instantaneous temperature defined by the average kinetic energy of the atoms. The corresponding averages over the entire trajectory are denoted by $\bar{X}_k$. We picked two independent snapshots for PI-DPMD and 10 snapshots for DPMD for pure water, which are the same as we did in our previous paper[29]. At the same time, to make sure, we picked the most representative solution structure of NaCl ions, we computed the oxygen coordination number distribution function to add the criteria for the ions solution structures for both PI-DPMD and DPMD, shown in Fig.2 (c) and (d). Near the minimum of the score function, we selected two independent snapshots for PI-DPMD, where their oxygen coordination number are 6 and 6 (in the first shell of $Na^+$ cation), 6.75 and 7.75 (in the first shell of $Cl^-$ anion), and ten independent snapshots for DPMD, whose distribution of the coordination number distribution is similar to the distribution of the total trajectory.

### B. Details of the GW-BSE XAS Calculations

As we adopted the same strategy as we reported in previous study[29], we will briefly report the procedure to perform the GW-BSE XAS calculation here. We first computed the mean-field wavefunctions as the starting point for our GW-BSE calculation using the DFT at the level of the generalized gradient approximations (GGA) of Perdew, Burke and Ernzerhof (PBE)[61], with Quantum ESPRESSO package[62]. The multiple-projector norm-conserving pseudopotentials that match the all-electrons potential for oxygen and hydrogen were generated by using the ONCVPSP package[63]. The DFT wavefunction was set to have a 200 Ry plane wave cutoff to converge the description of core electrons.

For the GW-BSE calculation, we used a modified version of the BerkeleyGW package[64–66]. The GW calculation was done only at the Γ point and we used a 20 Ry cutoff (with 10000 bands included in the sum over empty bands) for the plane-wave components of the dielectric matrix. This number has been proven to be sufficient to converge the bandgap of liquid water[67]. A $G_1W_0$ self-consistent calculation was firstly performed with the static COHSEX approximation in order to improve the quasiparticle wavefunction. In this step, we updated 160 occupied states, and 320 unoccupied states according to the off-diagonal matrix elements of the self-energy operator. We have proven[29] that these number of occupied states and unoccupied states were sufficient to describe the shape of XAS spectra in terms of the balance of computational cost and accuracy. After that, a standard one-shot $G_0W_0$ calculation was performed with 10000 bands (where the 160 occupied states and the first 320 unoccupied states are from self-consistent COHSEX), the frequency dependence in the dielectric matrix is captured using the Hybertsen-Louie generalized plasmon-pole model (HL-GPP)[64]. The BSE calculation was done with 29 core states for NaCl solution (32 core states for pure water) and 160 unoccupied states, which are enough to cover the energy range in which we are interested. For all the calculations, we solved the electron-hole excitations within the GW-BSE approach in the Tamm-Dancoff approximation (TDA). We used the same screening parameter for the screened exchange term of electron-hole interaction kernel to include the transitions between valence band states and continuum conduction band states[29].

After constructed the electron-hole interaction kernel, we obtained the optical transition matrix elements by diagonalizing the BSE Hamiltonian with the momentum operators[29].

Unlike the aforementioned frozen core hole methods[31–33,49–57], in which the position of the excited hole is known for each excited state, current GW-BSE approach treats the excited states in terms of the band structure, therefore, the position of excited electron/hole is unknown. To connect the excited state and position of excited electron/hole, we use the average electron/hole densities method to identify the position of excited electron and core hole for each exciton, in which the average electron and hole densities for each exciton $S$ are defined as: $\rho_e^S(\boldsymbol{r}_e) = \int |\Psi^S(\boldsymbol{r}_h, \boldsymbol{r}_e)|^2 d\boldsymbol{r}_h$ and $\rho_h^S(\boldsymbol{r}_h) = \int |\Psi^S(\boldsymbol{r}_h, \boldsymbol{r}_e)|^2 d\boldsymbol{r}_e$, respectively. This method is quite useful to analysis the excitons in pre-edge and main-edge regions, as they are more localized compared the ones in the post-edge region, we will discuss the details later.

### III. Results and Discussion

#### A. Theoretical XAS Spectra of NaCl Solution and Pure Water

We present our theoretical XAS spectra of NaCl solution and pure water at room temperature, as shown in Fig. 1(a), the corresponding experimental spectra are given for references[31,29]. As one can see, a good agreement is reached between theory and experiment in terms of spectra width and spectral intensity in Fig. 1(a). Especially, we reproduced the experimental observed pre-edge (~535 eV), main-edge (~538 eV), post-edge (~541 eV) features by using the rigorous treatment of electron-hole dynamics. Notably, compared to the XAS spectra of pure water, we can find a specific enhancement in the experimental XAS spectra of NaCl solution for the pre-edge, main-edge peaks, while a slight reduced feature for post-edge peak.

To confirm that our GW-BSE XAS calculation can capture these changes, here, we show the differential XAS spectra of NaCl solution and pure water as presented in Fig. 1(b). There is a good agreement between theory and experiment. Upon solvating NaCl ion pairs in pure water, the ideal tetrahedral H-bond network of water molecules within the first hydration shell undergoes distortion. Consequently, the exciton centered at the water molecule becomes more localized, leading to heightened prominence of both the pre-edge and main-edge features. In comparison, the change observed in the post-edge feature is less pronounced, suggesting that the long-range ordering remains relatively unaffected by the introduction of NaCl salt ions into the water. This finding aligns with the results from our previous study[27]. It is worth noting that the slight overestimation of the post-edge change in our theoretical spectra can be attributed to the limited cell size employed in our calculations. Due to the heavy computational cost to compute GW-BSE XAS spectra, the cell we used is 9.529 Å × 9.529 Å × 9.529 Å, this indicates that we only include the water molecules in the first two shells of ions. Therefore, the excitons in the post-edge region are strongly constrained in the space which leads to the lower level of agreement between simulation and experiment. This limitation can be addressed by employing a larger simulation cell that contains a greater number of ion pairs, thus mitigating the size effects. Nonetheless, the successful reproduction of these alterations induced by solvated ions attests to the reliability of our GW-BSE XAS approach as a theoretical tool for interpreting XAS experiments involving salt solutions.

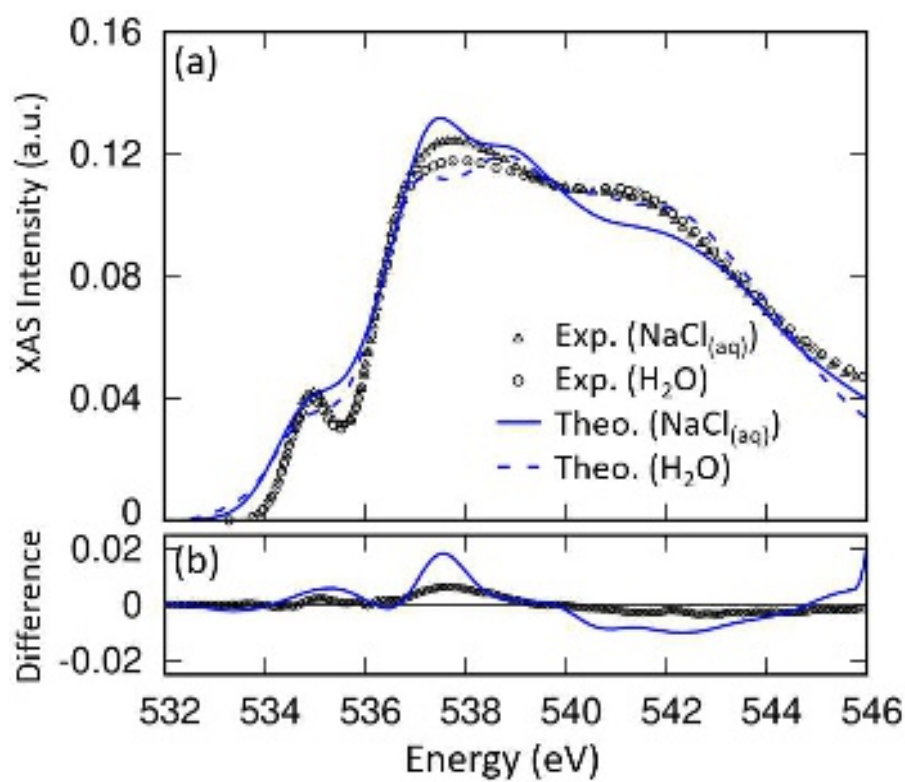

Fig. 1: (a) The experimental (Exp., black) and theoretical (Theo., blue) oxygen K-edge XAS of NaCl solution (triangles for Exp., solid line for Theo.) and liquid water (circles for Exp., broken lines for Theo.). The concentration for salt solution is corresponding to 2 M for both experimental and theoretical data. The experimental data is token from Ref. [33]. The theoretical XAS was generated at the GW-BSE level using the atomic configuration from a PI-DPMD simulation at 300K for both NaCl solution and pure water. The theoretical spectra are rigidly shifted by ~16 eV to align with the pre-edge of the experimental spectrum [29]. Both experimental and theoretical spectra are normalized in such a way that they have the same area within the same energy range from 532 eV to 546 eV. (b) Differential spectra of NaCl solution and liquid water: experiment (black circles) and theory (blue line).

## B. Impact of Nuclear Quantum Effects on the XAS Spectra

In this section, we will investigate the impact of NQEs on the XAS spectra of NaCl solution. The hydrogen is the lightest atom, whose NQEs cannot be neglected in water and aqueous solutions. Under the NQEs, there are two competing effects that either strengthen or weaken the H-bond[68]. The fluctuations of protons in the O-H covalent bond's stretching direction can aid in the formation of H-bonds. Conversely, the fluctuations in the proton's librational direction tend to weaken the strength of the H-bond network. It has been found that an approximately broadening effect has been induced by the delocalized protons, which slightly softens the water structure[35,43,46,69]. In the salt solution, there are three different interactions, water-cation, water-anion, and water-water, the latter two interactions contain two different H-bonds. Previously, we have reported[46] that NQEs are shown to weaken H-bonding between the chloride anion and the water molecules in the solvation shell, and the disruptive influence of the anion on the structure of the water solvent is notably reduced when NQEs are taken into account, in contrast to the scenario where NQEs are absent.

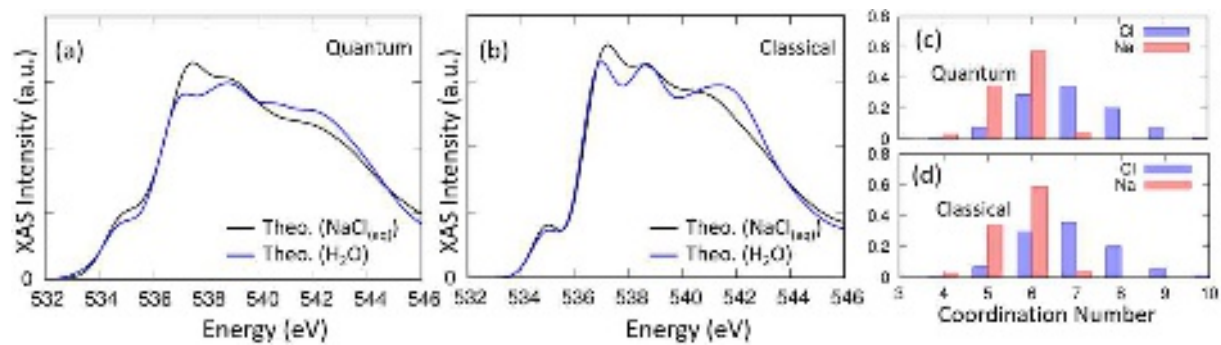

Fig. 2: The theoretical oxygen K-edge XAS of NaCl solution (black) and liquid water (blue) from PI-DPMD simulation (a) and DPMD (b). The coordination number of oxygen atom in the first hydration shell of $Cl^-$ (blue) and $Na^+$(red) ions in the classical (c) and quantum (d) simulation for NaCl solution. The radius of the first solvation shells of $Na^+$ is 3.2 Å for both classical and quantum simulations, while the radius of the first solvation shells of $Cl^-$ is 3.8 Å for both classical and quantum simulations.

Therefore, it is worth investigating how the XAS spectra of NaCl solution and water react to the presence of NQEs. To be more specific, how important is the NQEs in terms of the ion's effects on water molecules, does it change the nature of ions and water molecules? To answer such question, here, we show the theoretical XAS of NaCl solution and liquid water from PI-DPMD and DPMD simulation in Fig. 2 (a) and (b), respectively. Compared to the spectra obtained from the classical simulation, the spectra based on quantum simulation show a broader line shape, which is consistent with the previous reported PES spectra[43]. By including the NQEs, the theoretical spectra are closer to the experimental data. Moreover, the slight enhancement of the spectra obtained from quantum simulation compared with the one from classical simulation can be seen in the pre-edge region in both NaCl solution and pure water, it is a signature of the slightly more disordered proton configuration due to the NQEs. This is consistent with previous report[54]. Notably, we find compelling evidence that the absence of NQEs does not alter the characteristics of ions and their surrounding water molecules. This is evident from the enhanced pre-edge and main-edge peaks, along with the diminished post-edge peak observed in both quantum and classical scenarios. The rationale behind this observation is as follows: despite the weakening of the H-bond between the $Cl^-$ anion and water molecules caused by NQEs[46], the influence of solvating ions on the modification of molecular structures outweighs the impact of NQEs. NQEs can be considered as a perturbation to the interaction between ions and water molecules. Here, we show the coordination number of oxygen atom in the first hydration shell of $Cl^-$ and $Na^+$ ions in Fig. 2(c) and 2(d), one can find that they are sharing similar patterns. Therefore, we can discuss the impact of ions on the XAS spectra by theoretical spectra based on liquid water structures generated from classical DPMD simulations, as it takes less computational cost to calculate the XAS spectra and gives more information for the solvated structures.

### C. Influence of Ions on the Pre-edge of XAS Spectra

In this section, we will demonstrate how the solvating ions change the pre-edge feature of XAS spectra. The excitations associated with pre-edge feature of the XAS spectrum has been identified as the Frenkel-like excitons, which are sensitive to the short-range structural information about the H-bond network of water[22,70]. The electron-hole excitons in the pre-edge region have a strong intra-molecular character. The excited electron, with $4a_1$ symmetry is primarily located at the same molecule has the core hole[22,70]. In the experiment, the pre-edge peak is slightly enhanced in the spectra of the salt solution compared with the one from pure water. Previously, Saykally and coworkers have shown a correlation between the H-bond network for each water molecule in the first solvation shell of cations and the resulting XAS spectra[34]. Nevertheless, how the solvating ions change the electronic structure of water and result the enhancement of pre-edge peak is not clear. To examine it, we calculated the charge distribution of the excitons where the excited hole located in the first hydration shell of $Na^+$ and $Cl^-$ ions, compared them with the ones from pure water. The excitons within the pre-edge region are included. The radius of the first hydration shell of $Na^+$ and $Cl^-$ ions is defined as the position of the first minimum of the $g_{NaO}(r)$ and $g_{ClO}(r)$, which are 3.2 Å and 3.8 Å, respectively. The data is shown in Fig. 3(a). One can see that the clear enhancement of charge distribution at around 0.5 Å and 1.7 Å, which are corresponding to the charges of the excited $4a_1$ electron which located in oxygen and hydrogen atoms in the same molecule, respectively. This indicates that the enhancement of the pre-edge peak in the salt solution comes from the slightly localized excitons due to the solvated ions.

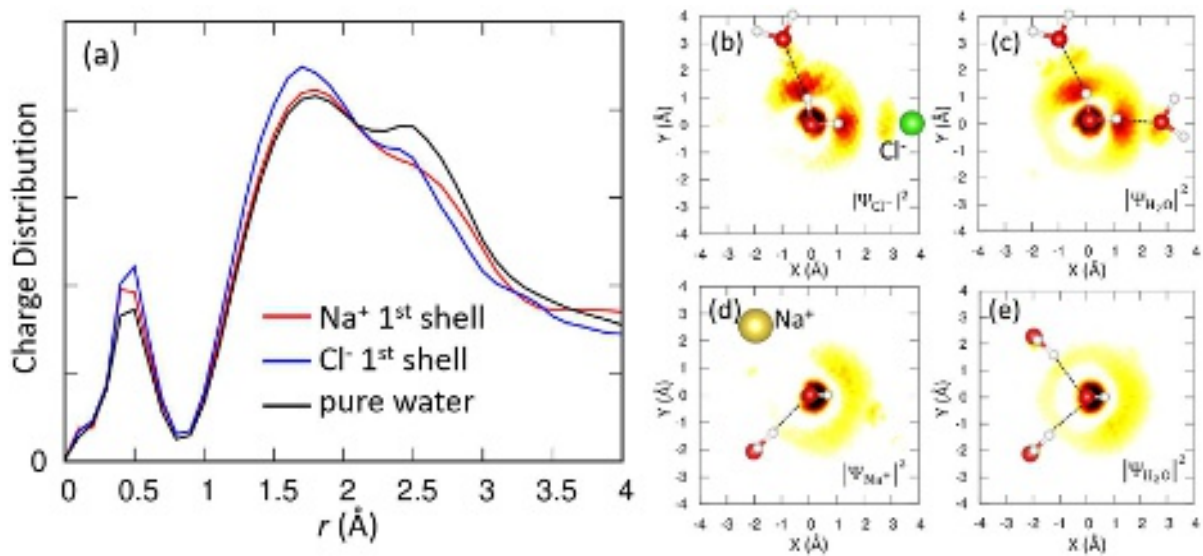

Fig. 3: (a) Charge distributions of the excitons in the pre-edge as a function of distance $r$ where the excited oxygen atoms are chosen from the ones within first shell of $Na^+$ (red), first shell of $Cl^-$ (blue), and from pure water (black). The zero point is fixed as the hole position. The two-dimensional contour plot of the electron density ($|\Psi|^2$) for excitons in the pre-edge peak. The cutting plane is defined as the plane (b-c) of $H_2O$ and the bisector plane of $H_2O$ (d-e) with a hole.

We now turn our attention to the detailed roles of $Na^+$ cation and $Cl^-$ anion in changing the localization of excitons in the pre-edge region. As we have argued that there are two different interactions between ions and water molecules: the $Cl^-$ anion is interacting with surrounding water molecules through H-bond, while the $Na^+$ cation has no H-bond interaction with water molecules. Here, we show the two-dimensional contour plot of the electron density $|\Psi_{Cl^-}|^2$ and $|\Psi_{H_2O}|^2$ for excitons in the pre-edge peak, where the hole located in the molecules within the 1$^{st}$ hydration shell of $Cl^-$ anion and pure water, in Fig. 3(b) and 3(c), respectively. The cutting plane is defined as the plane of $H_2O$ with a hole. The major difference between Fig. 3(c) and 3(d) is that asymmetric feature observed in the plot of $|\Psi_{Cl^-}|^2$, compared with the one of $|\Psi_{H_2O}|^2$. The perturbative effects of a Cl⁻ ion in a H-bond "acceptor" position, with respect to the central water molecule. The excited electron trends to be more localized along the water-water H-bond instead of Cl⁻-water H-bond directions, because of the gap between water and Cl anion in term of electronic structure. This is consistent with the findings of the less hybridization of the *p* orbitals between O atom and Cl atom observed in the PES spectra[43]. Therefore, one can expect a more localized exciton within the first hydration shell of Cl⁻ anion compared with the one from the pure water. Next, let us focus on the effects of Na⁺ cation on the pre-edge peak. Here, we also show the two-dimensional contour plot of the electron density $|\Psi_{Na^+}|^2$ and $|\Psi_{H_2O}|^2$ for excitons in the pre-edge peak, where the hole located in the molecules within the 1ˢᵗ hydration shell of Na⁺ anion and pure water, in Fig. 3(d) and 3(e), respectively. The cutting plane is defined as the bisector plane of H₂O with a hole. Instead of serving as a H-bond "acceptor," the Na⁺ cation is situated in the "donor" position. As one can see, the excited electron does not extend to its neighbor atoms due to the presence of Na⁺ cation. The resulting $4a_1$ excited electron is more localized along the direction of the donor H-bonds. Moreover, we can find the orbitals located on oxygen is slightly distorted, which reflects the presence of the positively charged Na⁺ cation. In short, the enhancement of the pre-edge peak can be explained as the presence of Cl⁻ anion and Na⁺ cation, which tend to localize the exciton electron wavefunction located at the same molecule has the core hole.

### D. Influence of Ions on the Main-edge of XAS Spectra

Now let us discuss the influences of ions on the main-edge of XAS spectra. Unlike the excitons in the pre-edge region, whose excited electrons are primarily located at the same molecule has the core hole, reflecting the information of the short-range H-bond network. The excitons in the main-edge peak region, are primarily categorized as intermolecular electron-hole excitations, as the excited electron is distributed on both the water molecule where the core hole is localized and the H-bonded water molecules in the coordination shell[22,70]. Let us reconcile the theory mentioned in the above, the influence of solvated ions is confined to the first hydration shell surrounding the ions, beyond which the H-bond network of water molecule remains similar as the ones in the pure water. If the aforementioned hypothesis holds true, it is anticipated that the presence of extra ions in the solution will alter the localizations of the main-edge excitons due to their influence on the H-bonding network of water molecules in the first hydration shell of ions, while the spectral contributions beyond the shell are likely to remain comparable to those of free water molecules.

First, we calculated the averaged intensity ratio between NaCl solution ($I_{\text{NaCl}}$) and pure water ($I_{\text{H}_2\text{O}}$) in the main-edge region: $I_{\text{NaCl}}/I_{\text{H}_2\text{O}} = 1.21$. Note that the excitons whose energies are within 1.5 eV window of main-edge peak are identified as the main-edge excitons. The corresponding spectra and absorption cross sections are shown in Fig. 4(a). Then, we decomposed the contributions of the excitons which are located within the first hydration shell of Na$^+$ ($I_{\text{Na}^+}$) and Cl$^-$ ($I_{\text{Cl}^-}$) ions, the ones outside these shells ($I_{\text{free}}$). The contributions within the first hydration shell of Na$^+$ ion is $I_{\text{Na}^+}/I_{\text{H}_2\text{O}} = 1.38$, the ones within the first hydration shell of Cl$^-$ anion is $I_{\text{Cl}^-}/I_{\text{H}_2\text{O}} = 1.48$, and the free water contribution is $I_{\text{free}}/I_{\text{H}_2\text{O}} = 1.01$. It is clear that the observed enhancement of the main-edge peak primarily come from the water molecules within the first hydration shell, while the water molecules outside the shell show similar as the ones in the pure water. This is consistent with the findings from the virtual scattering methods reported in the previous study[27].

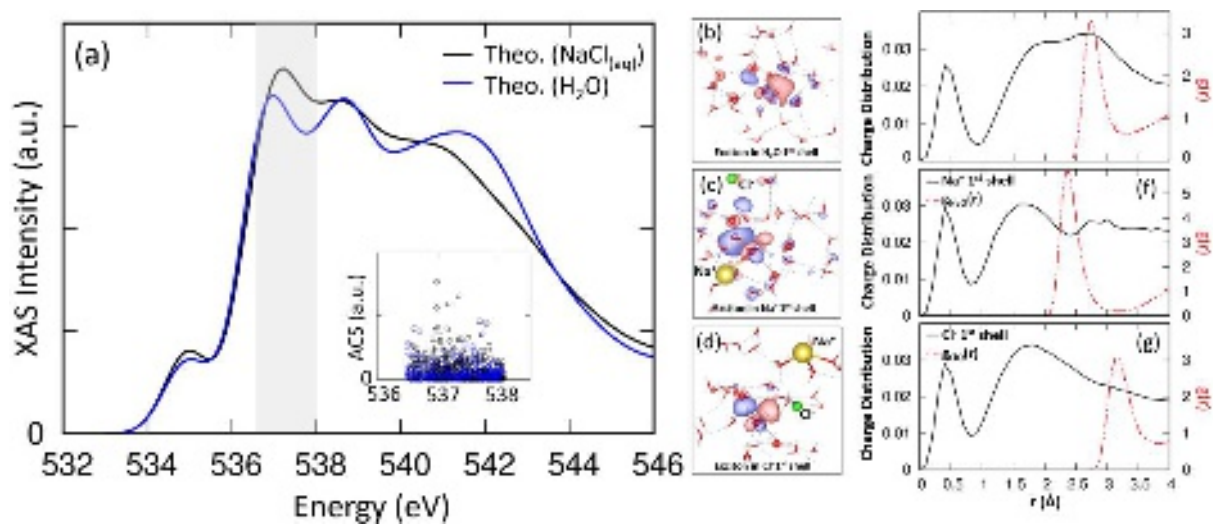

Fig. 4: (a) The theoretical oxygen K-edge XAS of NaCl solution (black) and liquid water (blue) from DPMD. The inserted figure is absorption cross sections (ACS) of excitation within main-edge window. The representative schematic of the excitons where the excited holes are placed at the oxygen atoms from pure water (b), and within first shell of Na$^+$ (c) and first shell of Cl$^-$ (d). The charge distributions (black line) of the excitons in the main-edge as a function of distance $r$ where the excited oxygen atoms are chosen from the ones from pure water (e), and within first shell of Na$^+$ (f) and first shell of Cl$^-$ (g). The radial distribution functions (red line) of $g_{\text{OO}}(r)$, $g_{\text{NaO}}(r)$, and $g_{\text{ClO}}(r)$ are shown for comparisons.

To further illustrate the effects of the ions altering the localizations of the excitons in the main-edge peak, we show the representative schematic as well as the charge distributions of the excitons where the excited holes are placed at the oxygen atoms from pure water and within the first shell of $Na^+$ and $Cl^-$ ions, in Fig. 4(b,e), 4(c,f) and 4(d,g), respectively. The radial distribution functions of $g_{OO}(r)$, $g_{NaO}(r)$, and $g_{ClO}(r)$ are shown for comparisons. In the pure water, the excited electrons of the main-edge exciton, with $b_2$ symmetry, are extended to the neighbor water molecules through H-bonds. These orbitals have larger lobes protruding from the covalently bonded hydrogens and have smaller lobes on the lone pair side of oxygen, which gives a peak feature at around 2 Å in Fig. 4 (e). The resonance states at the H-bond connecting water molecules give the peak feature at around 2.7 Å, shown in Fig.4(e), as it is the position of first peak of radial distribution function of $g_{OO}(r)$. Now let us focus on the $Na^+$ case, the $Na^+$ cation replaces the water molecules in the H-bond acceptor side with a shorter O-Na distance, and resonance exciton wavefunction won't be found at Na atom due to the interrupted H-bond network and the energy mismatch between Na and O. Therefore, the charge distribution shows a reduced amplitude at around 2.4 Å. The exciton orbitals of the molecules within the first hydration shell of $Na^+$ cation are more localized in the direction of the lone pair of the water molecule has the core hole, which contribute to a higher main-edge peak. After discussions of the effects of $Na^+$ cation, we discuss the effects of the $Cl^-$ anion. The $Cl^-$ anion forms a H-bond with the water molecules in the donor side, with a larger Cl-O distance. As such, we cannot find a peak in Fig. 4(g) at around 2.7 Å, which is existed in Fig. 4(e). The exciton wavefunctions within the first hydration shell of $Cl^-$ anion are more localized along the covalent O-H bond, compared with the ones in pure water, contribute to a higher main-edge peak. In summary, the effects of the solvating ions on the main-edge peak can be summarized as the ions intrude the water and disrupt the H-bond network so that the exciton wavefunctions would not be able to extend to the neighbor molecules due to the missing H-bonds.

### IV.  Conclusions

In summary, we have calculated the XAS spectra of NaCl solution by using the newly developed GW-BSE approach. Our theoretical XAS spectra of salt solution and pure water successfully reproduced the experimental observed the enhancement of the pre-edge, main-edge peaks, and a slight reduced feature for post-edge peak in the spectra of salt solution respect to the one of pure water. Our XAS results furthermore supports the theory that the influence of solvated ions is limited to the first hydration shell surrounding the ions, beyond which the H-bond network of water molecule remains similar as the ones in the pure water. We have provided detailed explanations for the spectral differences induced by the additional NaCl ions, which are typically attributed to the replacement effects of ion pairs within the tetrahedral H-bond network of water molecules. In the pre-edge region, the presence of additional ions distorts the wavefunctions of excitons located in the same water molecule as the core hole, leading to the enhancement of the pre-edge peak. Conversely, in the main-edge region, the resonance exciton wavefunctions are unable to extend to neighboring water molecules through the H-bond network due to the presence of ion pairs, resulting in a higher main-edge peak. We would like to note that despite utilizing the frozen core hole assumption, the results from previous studies[31–33] and our current work yield similar conclusions. Our rigorous treatment of the electron-hole interaction provides a nearly quantitative agreement between experiments and theory other than the qualitative agreement, revealing the effects of the ions on the molecular structures of the solution. In future work, it would be intriguing to explore whether our GW-BSE approach can accurately capture the delicate XAS features of various ice phases[30].

**Acknowledgments**

We thank Dr. Chunyi Zhang for helpful discussions. This work was supported by the Computational Chemical Center: Chemistry in Solution and at Interfaces funded by The DoE under Award No. DE-SC0019394. The work of K.S. was supported by National Science Foundation through Award No. DMR-2053195. The computational work used resources of the National Energy Research Scientific Computing Center (NERSC), a U.S. Department of Energy Office of Science User Facility operated under Contract No. DE-AC02-05CH11231. And this research includes calculations carried out on Temple University's HPC resources and thus was supported in part by the National Science Foundation through major research instrumentation grant number 1625061 and by the US Army Research Laboratory under contract number W911NF-16-2-0189.